\documentstyle[prb,aps,multicol,epsfig]{revtex}
\begin{document}

\title{The implications of resonant x-ray scattering data on the 
physics of the insulating phase of V$_{2}$O$_{3}$. }
\author{Y. Joly$^{1}$, S. Di Matteo$^{2,3}$, C. R. Natoli$^{2}$ }
\address {$^{1}${Laboratoire de cristallographie, CNRS, BP166,
F-38042 Grenoble Cedex 9, France}\\
$^{2}${Laboratori Nazionali di Frascati INFN, via E. Fermi 40 I-00044
Frascati (Roma) Italy}\\
$^{3}${Dip. Fisica "E. Amaldi", Universita di Roma 3, via della Vasca Navale 
84, I-00146 Roma Italy}\\}
\date{\today}
\maketitle
\draft
\begin{abstract}
We have performed a quantitative analysis of recent resonant x-ray scattering 
experiments carried out in the antiferromagnetic phase of V$_2$O$_3$ by means 
of numerical ab-initio simulations.
In order to treat magnetic effects, we have developed a method based on 
multiple scattering theory (MST) and a relativistic 
extension of the Schr\"{o}dinger Equation, thereby working 
with the usual non relativistic set of quantum numbers $l,m,\sigma$ for 
angular and spin momenta. 
Electric dipole-dipole (E1-E1), dipole-quadrupole (E1-E2) and 
quadrupole-quadrupole (E2-E2) transition were considered altogether.
We obtain satisfactory agreement with experiments, both in energy  
and azimuthal scans. All the main 
features of the V K edge Bragg-forbidden reflections with $h+k+l=$odd can be 
interpreted in terms of the antiferromagnetic ordering only, {\it ie}, they 
are of magnetic origin.  
In particular the ab-initio simulation of the energy scan around the 
(1,1,1)-monoclinic reflection excludes the possibility of any symmetry 
reduction due to a time-reversal breaking induced by orbital ordering.

\end{abstract}
\begin{multicols}{2} 

\section{Introduction}
In the last fifteen years, after the discovery of high-temperature 
superconductivity in cuprates, there has been an upsurge of renewed interest 
in the electronic properties of transition metal oxides.\cite{imada} Among 
the different techniques applied to investigate such materials, resonant 
x-ray scattering (RXS) has proved to be a powerful tool to extract direct 
informations about magnetic\cite{gibbs,hannon} and electronic
\cite{murakami,hirota} distributions: no other methods have the same 
flexibility to detect lattice, orbital and magnetic anisotropies within the 
same experimental setup.\cite{kcuf3} Yet, a general theoretical comprehension 
of some important implications of this technique is still lacking, and this 
circumstance has sometimes led to incorrect conclusions about the origin of 
the anomalous signals.\cite{maekawa,ezhov,mila} 
This is particularly true for magnetic RXS, where the absence of numerical 
{\it ab-initio} simulations has strongly limited the possibility of 
quantitative investigations. Only recently 
few papers have appeared dealing with this matter. In particular Takahashi 
{\it et al.}\cite{igarashi} have used the {\it ab initio} local density 
approximation (LDA) + $U$ scheme, taking also into account spin-orbit 
interaction, to describe such phenomena. In this way they were able to 
calculate the magnetic RXS in the energy region of the $4p$ conduction band 
in KCuF$_3$. 

One of the purposes of this paper is to present an alternative method to 
deal with magnetic phenomena, which we believe physically more intuitive, 
based on a relativistic extension of the Schr\"{o}dinger Equation and the 
multiple scattering theory (MST). This extension is obtained by eliminating 
the small component of the relativistic wave-function in the Dirac Equation 
and working only with the upper component,\cite{wood} thereby using the more
intuitive non relativistic set of quantum numbers $(l,m,\sigma)$ for 
the angular and spin momenta. In order to take into account spin-polarized 
potentials and spin-orbit interaction in the framework of MST we solve a 
two-channel problem, in each atomic sphere, for the two spin components of the 
wave-function coupled by the spin-orbit interaction. The condition 
$m + \sigma = m' + \sigma'$ must be satisfied, due to the local conservation 
of the 
$z$-projection of total angular momentum $\vec{j} = \vec{l} + \vec{\sigma}$. 
We have implemented the computer code to calculate magnetic RXS into the 
FDMNES package.\cite{yves,yvesfdmnes}

Secondly and more important, we present here the first application of this 
method to deal with the case of V$_2$O$_3$, which has attracted considerable 
attention in the past four years,
\cite{ezhov,mila,fabrizio,paolasini1,paolasini2,dimatteo,lovesey2,tanaka,lindic} due to the peculiar interplay of magnetic and orbital ordering. By 
performing  a quantitative analysis of the recent RXS experiments carried 
out by Paolasini {\it et al.}\cite{paolasini1,paolasini2} at the vanadium K 
edge, we shall try to establish the magnetic space group of the monoclinic 
phase, in the endeavor to resolve some controversies that are still going on 
in the scientific community. 

A brief history of the most recent findings about V$_2$O$_3$ will motivate 
our work. This compound is a Mott-Hubbard system showing a metal-insulator 
transition at around 150 K from a paramagnetic metallic (PM) to an
anti-ferromagnetic insulating (AFI) phase due to the interplay between band 
formation and electron Coulomb correlation.\cite{mcwhan} A structural phase 
transition takes place, at the same temperature, from corundum to monoclinic 
crystal class.
In recent years, on the basis of an old theoretical model by Castellani  
{\it et al.},\cite{cnr} Fabrizio {\it et al.}\cite{fabrizio} 
suggested that a direct observation of orbital ordering (OO) could have been 
possible by means of RXS. Soon after Paolasini {\it et al.}\cite{paolasini1} 
interpreted the forbidden (111) monoclinic Bragg-forbidden reflection as an 
evidence of such OO. All this gave rise to an intense debate
\cite{ezhov,mila,paolasini2,dimatteo,lovesey2,tanaka,lindic} whose main 
achievement was to prove the incorrectness of the old Castellani {\it et al.} 
model. In fact, non-resonant x-ray magnetic scattering experiments
\cite{paolasini1} and absorption linear dichroism\cite{park} showed that 
the atomic spin on vanadium ions is S=1, and not S=1/2, as supposed in 
Ref. [\onlinecite{cnr}]. 
Yet, none of the attempts to explain the origin of the (111) monoclinic 
reflection\cite{ezhov,mila,paolasini2,dimatteo,lovesey2,tanaka} can be 
considered satisfactory, for reasons that will be clearer in the following. 
Here we just recall that many different physical mechanisms have been 
proposed, from the antiferro-quadrupolar ordering of the 3$d$ orbitals,
\cite{ezhov} to the combined action of the ordered orbital and magnetic 
degrees of freedom;\cite{mila} from an orbital ordering associated with a 
reduction of the magnetic symmetry,\cite{dimatteo} to the anisotropies in the 
magnetic octupolar\cite{lovesey2} or toroidal distribution.\cite{tanaka}

In this paper we shall focus on a series of Bragg-forbidden reflections 
measured by Paolasini and collaborators.\cite{paolasini1,paolasini2} Of 
particular interest, from a theoretical point of view, is the set of data 
of Ref. [\onlinecite{paolasini2}], where the different reflections have 
been collected with particular care on their relative intensities, and 
energy and azimuthal scans have been performed in both the rotated 
($\sigma\pi$) and unrotated ($\sigma\sigma$) channels.
Thanks to the wide information content of these data, we are able to 
demonstrate that the signal cannot be due to any kind of orbital ordering 
or charge anisotropy and it must be magnetic. Moreover we can show that 
both mechanisms suggested in Refs. [\onlinecite{lovesey2,tanaka}] can be 
at work for the (1,1,1) reflection, as there is no extinction rule for either 
dipole-quadrupole or  quadrupole-quadrupole transitions. Which of the two 
contributions dominates depends strongly on the kind of reflection and on 
the azimuthal angle. To reach this conclusion {\it ab initio} calculations 
have proved necessary. 

The next section is devoted to the description of our method to deal 
with magnetic phenomena without resorting to the Dirac equation. We shall 
work in the framework of MST with muffin-tin approximation or within the 
finite difference method (FDM),\cite{yves} ie, without approximation on the 
geometrical shape of the potential.
In Sec. III we introduce the crystal and electronic properties of 
V$_2$O$_3$ and compare our results with the experimental data, discussing 
the physical mechanism behind each reflection. 
Finally we draw some conclusions on the possibility of gaining information 
about orbital and magnetic degrees of freedom by means of RXS.

\section{Method of calculation}

\subsection{RXS and related spectroscopies}

Core resonant spectroscopies are described by the virtual processes that 
promote a
core electron to some empty energy levels. They all depend on the transition 
matrix elements of the operator $\hat{O}$ expressing the interaction of 
electromagnetic radiation with matter:

\begin{equation}
M_{ng} = \langle \psi_n|\hat{O} |\psi_g \rangle
\label{eq_m}
\end{equation}

Here $\psi_g$ is the ground state and $\psi_n$ the photo-excited state.
In the x-ray regime, the operator $\hat{O}$ is usually written through the 
multipolar expansion of the photon field up to the electric quadrupole 
term\cite{note1}:

\begin{equation}
\hat{O}^{i(o)} = \vec{\epsilon}^{i(o)} \cdot \vec{r} ~ \big(1 - 
\frac{1}{2}i\vec{k}^{i(o)} \cdot \vec{r}\big)
\label{eq_2}
\end{equation}

\noindent where $\vec{r}$ is the electron position measured from the 
absorbing ion, $\vec{\epsilon}^{i(o)}$ is the polarization of the incoming 
(outgoing) photon and  $\vec{k}^{i(o)}$ its corresponding wave vector. In the
following dipole and quadrupole will refer to electric dipole and electric
quadrupole. In RXS the global process of photon absorption, virtual 
excitation of the photoelectron, and subsequent decay with re-emission of a 
photon, is coherent throughout the crystal. Such a coherence gives rise to 
the usual Bragg diffraction condition, that can be expressed, at resonance, as:

\begin{equation}
F = \sum_{a}e^{i\vec{Q}\cdot\vec{R}_a}(f_{0a}+f_a'+if_a'')
\label{eq_f}
\end{equation}

\noindent where $\vec{R}_a$ stands for the position of the scattering ion $a$,
$\vec{Q}$ is the diffraction vector and  $f_{0}$ is the usual Thomson factor. 
The resonant part, $f'+if''$, is the anomalous atomic scattering factor (ASF), given by the expression\cite{blume}:

\begin{equation}
f'+if'' = \frac{m_e}{\hbar^2}\frac{1}{\hbar\omega} \sum_{n}
\frac{(E_n-E_g)^3M_{ng}^{o*}M_{ng}^i}
{\hbar\omega-(E_n-E_g)-i\frac{\Gamma_n}{2}} \label{eq_arxs}
\end{equation}

\vspace{-0.2cm}

Here $\hbar\omega$ is the photon energy, $m_e$ the
electron mass, $E_g$ the ground state energy, and $E_n$ and $\Gamma_n$ are 
the energy and inverse lifetime of the excited states. 
In practice, the intermediate states $\psi_n$ are in the
continuum, normalized to one state per Rydberg. For this reason it is useful 
to label them by their energy $E$ and re-express Eq. (\ref{eq_arxs}) as:

\begin{eqnarray}
f'+if'' =  -m_e\omega^2 \sum_{n} \int_{E_F}^{\infty}
\frac{{M}_{ng}^{o*}{M}_{ng}^i}
{E-E_g-\hbar\omega-i\frac{\Gamma(E)}{2}}dE
\label{eq_arxs2}
\end{eqnarray}

\noindent using the fact that, at resonance, 
$E_n-E_g \simeq \hbar\omega$. $E_F$ is the Fermi energy. 
In Eq. (\ref{eq_arxs2}) the summation over $n$ is now limited to states 
having the energy $E$.

It is now straightforward to make the connection with another spectroscopic 
technique: x-ray near edge absorption (XANES).
In this case the cross-section simply corresponds
to the imaginary part of the ASF when $\hat{O}_o = \hat{O}_i$ 
(forward scattering). Choosing prefactors in order 
to have $f''$ in units of the classical electron radius, 
$r_0 \simeq 2.82 \cdot 10^{-15}$ m, and the absorption cross section
$\sigma$ in megabarn, we have the relation:

\begin{equation}
\sigma = - 4\pi~10^{22}~\frac{a_0^2\alpha^3m_ec^2}{\hbar\omega}f''
\label{eq_sigma}
\end{equation}

\noindent where $a_0$ is the Bohr radius, $\alpha$ is the
fine structure constant and $c$ is the speed of light. The
close connection between the two spectroscopies is thus evident.

The use of RXS can have some advantages in exploring electronic properties 
compared to the absorption techniques. First, with RXS it is possible to 
select different relative conditions in incoming and outgoing polarizations 
and wave vectors, and this gives more opportunities to probe 
the magnetic and electronic anisotropies of the material. Secondly, RXS can 
be more site selective than XANES, due to the Bragg factor (\ref{eq_f}).
For instance, it is possible, in magnetite, to be sensitive only to 
octahedral Fe$^{3+}$-sites and not to tetrahedral Fe$^{3.5+}$-sites
\cite{garcia2} or, in manganites, to probe only Mn$^{3+}$-ions and not 
Mn$^{4+}$-ions.\cite{tapan} This can be achieved at reflections that are 
Bragg-forbidden off-resonance, but become detectable, in resonant conditions, 
because of non-symmorphic symmetry elements (glide planes, screw axes) in the 
crystal space group.\cite{dmitrienko} In these cases the Thomson factors 
$f_{0}$ (Eq. \ref{eq_f}) drop out and one is just sensitive to the changes 
of the ASF (Eq. (\ref{eq_arxs})) that depend on the relative electronic and 
magnetic anisotropies on the ions related by such non-symmorphic symmetry 
elements. It is indeed in such kinds of reflections that orbital ordering or 
magnetic scattering were reported and they will be the main subject of the 
present study.

\subsection{Cartesian Tensor approach}

In Sec. III we shall deal with the symmetry operations of the crystal
space group of V$_2$O$_3$, in order to evaluate the structure factor for RXS. 
Because of this, it is very useful to write explicitly the dipole
and quadrupole components of the matrix elements in Eq. (\ref{eq_arxs2}). 
Remembering that incoming and outgoing
x-rays can have different
polarizations and wave vectors, we get:

\begin{eqnarray}
\sum_{n} {M}_{ng}^{o*}{M}_{ng}^i & = & 
\sum_{\alpha\beta}\epsilon_{\alpha}^{o*}
\epsilon_{\beta}^i D_{\alpha\beta} \nonumber \\
 & - & \frac{i}{2} \sum_{\alpha\beta\gamma}
\epsilon_{\alpha}^{o*}\epsilon_{\beta}^i
(k_{\gamma}^iI_{\alpha\beta\gamma}-k_{\gamma}^oI_{\beta\alpha\gamma}^*)
\nonumber \\
 & + &\frac{1}{4} \sum_{\alpha\beta\gamma\delta}\epsilon_{\alpha}^{o*}
\epsilon_{\beta}^i
k_{\gamma}^ok_{\delta}^iQ_{\alpha\beta\gamma\delta}
\label{eq_tensor}
\end{eqnarray}

\noindent  where $\alpha$, $\beta$, $\gamma$ and $\delta$ are cartesian
coordinate labels and $D_{\alpha\beta}$,
$I_{\alpha\beta\gamma}$
 and $Q_{\alpha\beta\gamma\delta}$, the dipole-dipole (dd), dipole-quadrupole (dq) and quadrupole-quadrupole (qq) contributions, respectively.
Their explicit expression is given by:

\begin{eqnarray}
 D_{\alpha\beta}& =&  \sum_{n} \langle \psi_g|r_{\alpha}|\psi_n\rangle
  \langle \psi_n|r_{\beta}|\psi_g\rangle \nonumber \\
 I_{\alpha\beta\gamma}& =& \sum_{n} \langle \psi_g|r_{\alpha}|\psi_n\rangle
  \langle \psi_n|r_{\beta}r_{\gamma}|\psi_g\rangle  \nonumber \\
 Q_{\alpha\beta\gamma\delta}& =&  \sum_{n}
 \langle \psi_g|r_{\alpha}r_{\beta}|\psi_n\rangle
  \langle \psi_n|r_{\gamma}r_{\delta}|\psi_g\rangle
\label{eq_tensor2}
\end{eqnarray}

\noindent  Note that at a K-edge, in absence of spin-orbit interaction and 
time-reversal breaking potentials, $D$, $I$ and $Q$ are all real.

\subsection{Calculation of the excited states}

The main difficulty in the {\it ab-initio} evaluation of the ASF is the 
determination of the excited states. Two different procedures, MST and FDM, 
have already been developed for non-magnetic cases, and the corresponding 
packages\cite{yvesfdmnes} used in some XANES\cite{yvestio2} and
RXS\cite{benfatto,garcia} applications. 
In the following we introduce a relativistic extension of the {\it ab initio} 
cluster approach, for both MST and FDM, that is still
mono-electronic but includes the spin-orbit interaction and allows to handle 
magnetic processes.
The importance of the FDM
procedure to avoid the muffin-tin approximation has
already been reported\cite{yves} and will not be detailed here again.
Note only that with FDM the intermediate states are calculated by
solving the Schr\"odinger equation without any
approximation on the geometrical shape of the potential, ie, we do not need to 
use, as usually in MST approaches, a spherically averaged potential in the 
atomic spheres and a constant among them. This point is essential to study 
orbitally ordered materials, where an anisotropic orbital distribution is 
the key-feature of the system and no spherically averaged potentials would be 
adequate for its description. Unfortunately,
large cluster calculations with FDM are prohibitively long in computing time 
and require a lot of computer memory: that's why, when appropriate, MST is 
preferred.

In order to treat magnetic effects in MST we have considered the 
relativistic extension of Schr\"{o}dinger equation obtained 
by solving Dirac equation exactly for the upper component of the 
wave-function, as described by Wood and 
Boring.\cite{wood} This approach ensures that the singularity of 
the potential at the nuclear center is of centrifugal type 
($ \sim r^{-2} $) even in the presence of spin-orbit terms (so that the usual 
fuchsian method of solution can be used). It also allows us to 
work with the usual non relativistic set of quantum numbers ($l,m,\sigma$) 
of angular and spin momenta, which we find much more intuitive than the 
relativistic set.
Our one particle basis will therefore comprise spin-orbit coupled core states 
of the form:
\begin{eqnarray}
\label{boh1}
\Phi^c_{j,j_z;l_c}(\vec{r})& = & R_j^{c}(r) |j,j_z;l_c) \\
& = & R_j^{c}(r) \sum_{m_c\sigma} 
Y_{l_cm_c}\chi_{\sigma}\left( l_cm_c 1/2\sigma|j,j_z \right)
\nonumber
\end{eqnarray} 
and valence excited states which are multiple scattering solutions of the 
Schr\"{o}dinger 
equation with spin-polarized potentials. In atomic units:  

\begin{eqnarray*}
\left\{ \bigtriangledown^2 + k_e^2 - V_0(\vec{r}) - V_1(\vec{r})s_z - 
2V_2(\vec{r}) \vec{\ell} \cdot \vec{s} \right\} 
\Psi_{\vec{k}_e,s}(\vec{r}) = 0
\end{eqnarray*}

\vspace{-0.8cm}

\begin{equation}
\label{e0}
\end{equation}

\vspace{-0.3cm}

These latters are supplemented by the scattered wave boundary conditions:
\begin{eqnarray}
\Psi_{\vec{k}_e,s}(\vec{r}) = e^{i \vec{k}_e \cdot \vec{r}} 
\chi_s - f(\vec{k}_e,s; \vec{r},s') \frac{e^{k_e r}}{r} \chi_{s'}
\label{e00}
\end{eqnarray}

In Eq. (\ref{boh1}) $\left( l_cm_c 1/2\sigma|j,j_z \right)$ are the 
Clebsch-Gordan coefficients.
In Eq. (\ref{e0}) $V_0(\vec{r})$ is the average of spin up and spin 
down potentials,
$V_1(\vec{r})$ their difference and $V_2(\vec{r})$ the usual spin-orbit term.
Such potentials already embody the relativistic corrections 
due to the reduction of the upper Dirac component of the wave function to an 
effective Schr\"{o}dinger equation.\cite{wood}

In the muffin-tin approximation the solution inside the
 \emph{i-th} atomic muffin-tin sphere can be written as:
\begin{eqnarray*}
\Psi_{\vec{k}_e,s}(\vec{r}_i) = \sum_l \sum_{m\sigma} \sum_{m'\sigma'}   
R_{lm\sigma}^{lm'\sigma'}(r_i) B^i_{lm'\sigma'}(\vec{k}_e,s)
Y_{lm}(\hat{{r}}_i) \chi_{\sigma}
\end{eqnarray*}

\vspace{-0.8cm}

\begin{equation}
\label{e1}
\end{equation}

\vspace{-0.2cm}

\noindent where at the muffin-tin radius $\rho_i$ the radial functions 
$R_{lm\sigma}^{lm'\sigma'}(r_i)$ match smoothly to the following combination 
of Bessel and Hankel functions via the atomic $t_{lm\sigma, lm'\sigma'}$ 
matrices (defined below):
\begin{eqnarray}
R_{lm\sigma}^{lm'\sigma'}(\rho_i) = j_l(k\rho_i)
t^{-1}_{lm\sigma, lm'\sigma'}
-i\delta_{lm\sigma, lm'\sigma'}h_l^+(k\rho_i)
\end{eqnarray}

With a proper normalization of these radial functions 
to one state per Rydberg, the scattering amplitudes 
$ B^i_{L\sigma}(\vec{k}_e,s)$ obey the equations (writing $L$ for $lm$):
\begin{eqnarray*}
\sum_j \sum_{L'\sigma'} M^{ij}_{L\sigma,L'\sigma'}
B^j_{L'\sigma'}(\vec{k}_e,s) = 
\delta_{s\sigma} i^l Y_L(\hat{{k}}_e) 
e^{i \vec{k}_e \cdot \vec{R}_{i} }
\end{eqnarray*}
where 
$$M^{ij}_{L\sigma,L'\sigma'} =  
\left( t^{-1}_{lm\sigma,l'm'\sigma'} \delta_{ij} \delta_{ll'}
\delta_{m+\sigma,m'+\sigma'} + G^{ij}_{L,L'} \delta_{\sigma,\sigma'} \right)$$ 
is the usual multiple scattering matrix, generalized to spin variables, and 
$\vec{R}_i$ denotes the position of the \emph{i-th} atom in the 
cluster with respect to the origin of the coordinates. 
The atomic $t$-matrix $t_{lm\sigma, lm'\sigma'}$ describes the  
scattering amplitude of an electron impinging the atomic potential with 
angular momentum $l$, azimuthal component $m$ and spin $\sigma$ into a state 
with quantum numbers $l,m',\sigma'$.
The conservation of the total angular momentum
$\vec{j} = \vec{l} + \vec{s}$ and its $z$-projection implies the constraint 
$m+\sigma = m'+\sigma'$. Note that also $l$ is unchanged in the scattering 
process, since in the muffin-tin approximation the potential 
has spherical symmetry.

By introducing as usual\cite{dimatoli} the scattering path operator 
$\tau^{ij}_{L\sigma,L'\sigma'}$ as the inverse of 
$M^{ij}_{L\sigma,L'\sigma'}$,
the solution for the scattering amplitudes $B^i_{L\sigma}$ is given by:
\begin{eqnarray}
B^i_{L\sigma}(\vec{k}_e,s) = \delta_{\sigma s} \sum_j \sum_{L'} 
\tau^{ij}_{L\sigma,L's} i^{l'} Y_{L'}(\hat{{k}}_e) 
e^{i \vec{k}_e \cdot \vec{R}_j}
\label{e2}
\end{eqnarray}
The scattering path operator $ \tau^{ij}_{L\sigma,L'\sigma'} $ represents the 
probability amplitude
for the excited photoelectron to propagate from site $i$, with 
angular momentum $L$ and spin $\sigma$, to site 
$j$ with angular momentum $L'$ and spin $\sigma'$. It is the obvious 
generalization of the corresponding spin-independent quantity.\cite{dimatoli}

We now have all the ingredients to perform the intermediate sum in 
Eq. (\ref{eq_arxs2}). Using the expression given Eq. (\ref{e1}) for the 
state $|\psi_n \rangle$  in the matrix elements of the numerator, 
the sum over $n$ becomes an integral over the escape 
direction of the photoelectron, $\hat{k}_e$, at fixed energy, $E = k_e^2$. 
When we perform the integral over the energy, we get the product of two 
smoothly varying radial matrix elements of the type 
$\langle R_{lm\sigma}^{lm'\sigma'}(r)|r^{\nu}|R_j^c(r) \rangle$ times the 
appropriate angular (Gaunt) coefficient. The exponent ${\nu}$ is determined 
by the polar order of the transition. This is further multiplied by the 
angular integral of the 
product of the two scattering amplitudes $B^i_{L\sigma}$, $B^j_{L'\sigma'}$ 
centered on the 
absorbing sites. This latter can be simplified 
by the use of a generalized optical theorem\cite{Natoli86}: 
\begin{eqnarray}
\sum_s \int d{\hat{k}}_e B^i_{L\sigma}(\vec{k}_e,s) 
B^j_{L'\sigma'}(\vec{k}_e,s) = \Im \tau^{ij}_{L\sigma,L'\sigma'}
\label{e3}
\end{eqnarray}
As a consequence, the knowledge of $\tau$ at all relevant energies, and of the 
radial solutions $R_{lm\sigma}^{lm'\sigma'}(r)$ on the photoabsorber are 
sufficient to calculate the ASF.

In the case of V$_{2}$O$_{3}$, in order to construct the spin-polarized 
atomic potentials, we have used the prescription by von Barth and 
Hedin\cite{vbarth}: they were derived from the non-self-consistent spin-polarized charge density obtained by superimposing the
atomic charge densities with two magnetic electrons on each Vanadium ion, as 
suggested by the experimental data.\cite{paolasini1,park}
We believe that this approximation is not far from reality, as shown 
{\it a posteriori} by 
the goodness of the results. Given the potential, it is straightforward to 
calculate the atomic spin dependent $t$-matrix: we just solve the two-channel 
problem arising from the Schr\"{o}dinger equation (\ref{e0}) due to the 
presence of the spin-orbit potential. In this way this latter is treated 
on the same footing as the other potentials and not in perturbation theory 
as usually done.

The relativistic extension in the FDM scheme follows closely the previous 
treatment for MST. The space is partioned, as in the multiple scattering 
approach, in three regions\cite{yvesfdmnes}: {\it i)} An outer sphere 
surrounding the cluster where one impose the scattering behaviour of the 
wave function given in Eq. (\ref{e00}). {\it ii)} An 
atomic region made up of spheres, around each atom, with radii much smaller 
than the muffin-tin radii (of the order of 1 a.u. or less). Here the charge 
density is to a very good approximation spherically symmetric, due to the 
presence of the core electrons. {\it iii)} Finally, an interstitial region 
where the Laplacian operator of the Schr\"{o}dinger equation (\ref{e0}) is 
discretized and the solution is generated on a greed without any approximation 
on the geometrical shape of the potential. By imposing a smooth continuity of 
the overall wave function across the boundaries of the three regions, one can 
determine both the expansion amplitudes of the wave function inside each atom 
and the scattering T-matrix of the whole cluster in Eq. (\ref{e00}).
\cite{yves} 
Note that in this case the spin-orbit potential is not spherically symmetric. 
We have checked that for a small cluster with close-packed geometry, this 
approach and the multiple scattering approach with muffin-tin approximation 
provide almost 
identical cross sections. Since V$_{2}$O$_{3}$ is a close-packed structure,  
if one neglects vanadium voids, we have used MST in the 
muffin-tin approximation for the magnetic calculations. The FDM scheme 
was instead used in the calculation of the ASF in the case of orbital 
ordering.

\section{The case of V$_2$O$_3$}

In this section we specialize to the case of V$_2$O$_3$, to investigate 
whether the RXS reflections observed by Paolasini and collaborators
\cite{paolasini1,paolasini2} are of magnetic or orbital nature. 
Our result is that magnetic ordering can explain the experimental data with 
a reasonably good agreement (even if not perfect, for reasons that will be 
discussed below). On the other hand, we are able to demonstrate that any OO 
origin of the signal has to be excluded, as it would not fit the energy scan 
at dipolar energies.

\begin{figure}
\centerline{\epsfig{file=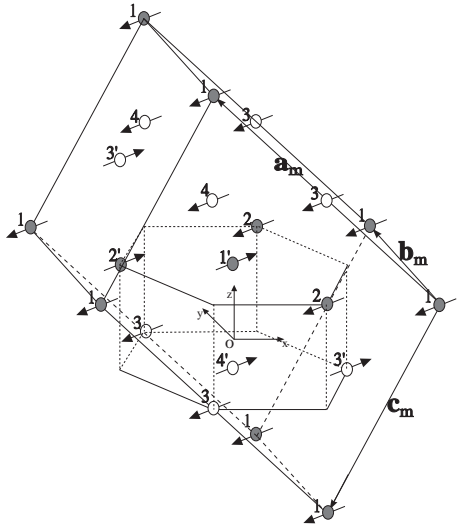,width=6cm}}
\vspace{0.7cm}
\caption {Monoclinic cell in the AFI
phase of V$_2$O$_3$. The numbers label the eight vanadium ions of the unit
cell. The arrows show the spin orientations. $a_m$, $b_m$ and $c_m$ are
the monoclinic axis.
 Only V ions are
 represented: white and gray circles indicate that they
are surrounded by differently oriented oxygen octahedra.
Two basal planes of the hexagonal corundum
cell are also shown.}
\label{fig_1}
\end{figure}

Next subsection is devoted to the description of the magnetic space group in 
the AFI phase, as well as the derivation of the anomalous RXS amplitude. In 
subsection B we analyze the {\it ab-initio} calculations for the magnetic 
RXS, through the procedure previously developed. Finally, in the last part 
we show that the consequences of a time-reversal breaking OO are not 
compatible with the experimental data, thus putting a severe constraint on 
the possible OO theories for V$_2$O$_3$.

\subsection{Crystal and magnetic structure}

As a first step we shall try to establish the magnetic symmetry group in 
the AFI phase. Our starting point is the crystallographic structure, as 
given by Dernier and Marezio\cite{dernier} together with the magnetic 
data given by Moon\cite{moon} and later confirmed by Wei Bao {\it et al.} 
\cite{weibao}

Referring to the frame and the numbering of
vanadium atoms of Fig. \ref{fig_1}, we can divide the eight atoms
of the monoclinic unit cell into two groups of four, $V_1, V_2, V_3, V_4$ and
$V_{1'}$, $V_{2'}$, $V_{3'}$, $V_{4'}$, with opposite
orientations of the magnetic moment. The two groups are related by a 
body-centered translation.
Vanadium-ion magnetic
moments, indicated by arrows in the figure, lie
perpendicular to the $b_m$-axis at an angle of 138$^0$ away from the
 monoclinic $c_m$-axis. Neglecting the magnetic moments, the previous two
groups of four atoms with their oxygen environments are translationally
equivalent.

We can infer from these
data that the magnetic space group can be written as
$P2/a+{\hat {T}}\{{\hat{E}}|t_0\}P2/a$, where ${\hat {T}}$ is the time-reversal
operator and $t_0$ the body-centered translation. The monoclinic group $P2/a$ contains four symmetry operations:
the identity ${\hat {E}}$, the inversion ${\hat {I}}$, the two-fold rotation
about the monoclinic $b_m$ axis ${\hat {C}}_{2b}$ and the reflection
${\hat {m}}_b$ with respect to the plane perpendicular to this axis.
All these operations are associated to the appropriate translations, as shown in the following table:

\begin{eqnarray*}
\begin{array}{c|cccc}
 {\rm atom} &  {\rm position} & {\rm spin} & {\rm sym.} & {\rm translation } \\
\hline
V_1  &  (u,v,w)    & \uparrow  &  {\hat E}   & 0   \\
V_2  &  (-u,-v,-w) & \uparrow  &  {\hat I}   & 0   \\
V_3  &  (\frac{1}{2}+u,-v,w) & \uparrow  & {\hat T \hat m_b} &
              (\frac{1}{2},0,0)        \\
V_4  &(\frac{1}{2}-u,v,-w)   & \uparrow  & {\hat T \hat C_{2b}} &
              (\frac{1}{2},0,0) \\
V_{1'}  & (\frac{1}{2}+u,\frac{1}{2}+v,\frac{1}{2}+w) & \downarrow & {\hat T} &
                   (\frac{1}{2},\frac{1}{2},\frac{1}{2}) \\
V_{2'}  & (\frac{1}{2}-u,\frac{1}{2}-v,\frac{1}{2}-w) & \downarrow &
             {\hat T \hat I} & (\frac{1}{2},\frac{1}{2},\frac{1}{2})  \\
V_{3'}  & (u,\frac{1}{2}-v,\frac{1}{2}+w) & \downarrow &
             {\hat m_b } & (0,\frac{1}{2},\frac{1}{2}) \\
V_{4'}  & (-u,\frac{1}{2}+v,\frac{1}{2}-w) & \downarrow &
             {\hat  C_{2b}} & (0,\frac{1}{2},\frac{1}{2})
\end{array}
\end{eqnarray*}
\begin{center}
TABLE I.
\end{center}

Here $u = 0.3438, \; v = 0.0008, \; w = 0.2991$ are the fractional
coordinates of the atoms in unit of the monoclinic axis. Notice that it is 
sufficient to calculate the ASF for atom number 1, since all the others can 
be deduced by applying the symmetry operation indicated in the fourth column 
of the table.

In the following, we re-analyze the extinction rules for
the different reflections $(h,k,l)$ measured by Paolasini {\it et al.}
\cite{paolasini1,paolasini2}
in terms of the crystal tensors in the dd, dq and qq channels 
(Eq. (\ref{eq_tensor2})). All the recorded
reflections have the same incoming polarization, $\sigma$  
(electric field perpendicular to the diffraction plane), whereas the 
outgoing polarization is analyzed both in the $\sigma$  and $\pi$  channels
 (this latter with the electric field in the diffraction plane). 
The azimuthal scans are then registered by
rotating the sample around the diffraction vector. We refer to Paolasini 
{\it et al.}
[\onlinecite{paolasini2}] for the definition
of the azimuthal origin: the situation where the scattering
plane contains $a$ and $c$ hexagonal axes corresponds to an azimuthal 
angle of 90 degrees.

From equation (\ref{eq_f}) and Table I, 
neglecting the very small component of $\vec{R}$ along $b_m$, ($v=0.0008$), 
we get for each reflection $(h,k,l)$ the expression:

\begin{eqnarray}
{\cal F}^{hkl}=(1+(-)^h\hat{T}\hat{m_b}) (e^{i\phi_{hkl}}+e^{-i\phi_{hkl}}
\hat{I}) \nonumber \\
(1+(-)^{h+k+l}\hat{T})f_1
\label{eq_t}
\end{eqnarray}

\noindent where $R\simeq (u,0,w)$ is the position of the $V_1$ atom, 
$\phi_{hkl}\equiv \vec{Q}\cdot\vec{R}\equiv 2\pi (hu+lw)$ and $f_1$ stands 
for the ASF as defined in (Eq. \ref{eq_arxs2}). Note that $f_1$ is a scalar 
and we adopt the convention, when we say that a symmetry operator acts on 
$f_1$, that it acts only on the tensor components (dd, dq and qq) given by 
Eq. (\ref{eq_tensor2}).
Focusing on the reflections with $h+k+l=$odd and noticing that 
$D_{\alpha\beta}$ and $Q_{\alpha\beta\gamma\delta}$ are inversion-even, while
$I_{\alpha\beta\gamma}$ is inversion-odd, we find that only the following 
tensor components contribute to the signal:

\begin{eqnarray}
{\cal D}_{\alpha\beta}^{hkl} = 4i(1+(-1)^{h+n_y+1})\cos(\phi_{hkl})
\Im(D_{\alpha\beta}) \nonumber \\
{\cal I}_{\alpha\beta\gamma}^{hkl} = 4(1+(-1)^{h+n_y+1})
\sin(\phi_{hkl})\Im(I_{\alpha\beta\gamma}) \nonumber \\
{\cal Q}_{\alpha\beta\gamma\delta}^{hkl} = 4i(1+(-1)^{h+n_y+1})
\cos(\phi_{hkl})\Im(Q_{\alpha\beta\gamma\delta})
\label{eq_d}
\end{eqnarray}

\noindent where $n_y$ is the number of $y$ labels among the tensor indices 
$\alpha$, $\beta$,
$\gamma$, and $\delta$. Thus, in order to have a signal, 
$h$ and $n_y$ must have different parity. Notice that all
the quantities in Eq. (\ref{eq_d}) are magnetic, as only imaginary parts of 
cartesian tensors are involved. Moreover, the three scattering amplitudes are 
all purely imaginary, as the dq polarizations carry an extra imaginary unit 
(see Eq. \ref{eq_tensor}). As a consequence, they all interfere.
A separate analysis of the three tensors of Eq. (\ref{eq_d}) gives the 
following indications.

In the case of dd tensors, when $h$ is odd, the only non-zero component is
${\cal D}_{xz}^{hkl}-{\cal D}_{zx}^{hkl}\propto \langle L_y \rangle$.
\cite{nota0} When $h$ is even the non-zero components are 
${\cal D}_{xy}^{hkl}-{\cal D}_{yx}^{hkl}\propto \langle L_z \rangle$ and 
${\cal D}_{yz}^{hkl}-{\cal D}_{zy}^{hkl}\propto \langle L_x \rangle$. 
As the magnetic moment direction is perpendicular to
the $b_m$-axis, $\langle L_y \rangle$ is zero. Thus for odd $h$ no signal 
is expected in the dipolar region. On the contrary, when $h$ is even, a dd 
contribution is present. 
Of course, no $\sigma\sigma$ magnetic scattering is allowed.\cite{blume}
These facts explain why the experimental
spectrum for the $(2,\bar{2},1)_{\sigma\pi}$ shows structures at
the $4p$ edge, contrary to the $(2,\bar{2},1)_{\sigma\sigma}$
reflection and to the $(1,1,1)$ and $(3,\bar{1},1)$ reflections for
both polarization conditions. 
Notice that these results had been already derived with similar methods.
\cite{dimatteo,lovesey2,tanaka}

These remarks do not apply for qq and dq tensors. They always
contribute, for all the investigated reflections, in both
$\sigma\sigma$ and $\sigma\pi$ channels, except at some very specific 
azimuthal angles.
At the K edge these tensors measure the transitions to the
states with pure 3$d$ or hybridized
$3d$-$4p$ character. For this reason they have a finite value just close to 
the Fermi energy.
The physical quantities measured by these operators can be identified as the 
octupolar magnetic moment for the qq term and the toroidal or quadrupolar 
magnetic moment for the dq tensor.

\vspace{-20pt}

\subsection{Analysis of the magnetic signal}

In order to get the absolute intensities
and shape of the spectra, we need now to resort to {\it ab initio} 
calculations.
We performed such calculations for the crystal and magnetic
structures given in Table I. Thus, we do not neglect the small $v$ value
but we shall see that the conclusion given in the previous subsection
is not modified. The potential is calculated using a
superposition of atomic densities obtained from an atomic,
self-consistent Hartree-Fock calculation, with a $3d^2$, spin 1,
configuration. The MST approach is used with different
cluster radii from 3.0 up to 7.2 {\AA}, ie,  from a $VO_6$
molecule to a cluster containing 153 atoms.

\begin{figure}
 \vspace{40pt}
 \centerline{\epsfig{file=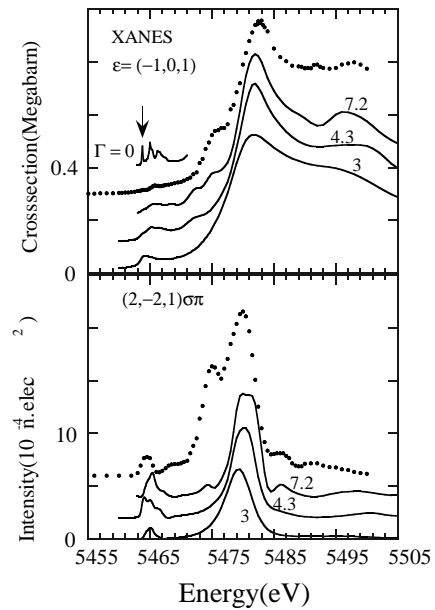,width=6cm}}
 \vspace{30pt}
 \caption {Comparison of
the experimental (dots) and calculated (continuous line) spectra
for the XANES and the $(2,\bar{2},1)_{\sigma\pi}$ reflection at the vanadium K
edge. Calculations are performed with three different cluster radii (3, 4.3 
and 7.2 {\AA}).
  The pre-edge feature is also shown 
 before the convolution for the 4.3 {\AA} case. The first structure, 
indicated by an arrow,
 is at the energy where the 3d feature of the Bragg peaks appears.}
\label{figure3_conv}
\end{figure}

We find that, in order to get the shoulder at 5475
eV in both XANES and $(2,\bar{2},1)_{\sigma\pi}$ spectra, as well as the
 5487 eV shoulder in the $(2,\bar{2},1)_{\sigma\pi}$, we need the biggest 
cluster radius (7.2 {\AA} - see Fig. \ref{figure3_conv}). Nevertheless the 
main features are present for all Bragg peaks even for the $VO_6$ molecule 
calculation.

\begin{figure}
 \vspace{40pt}
 \centerline{\epsfig{file=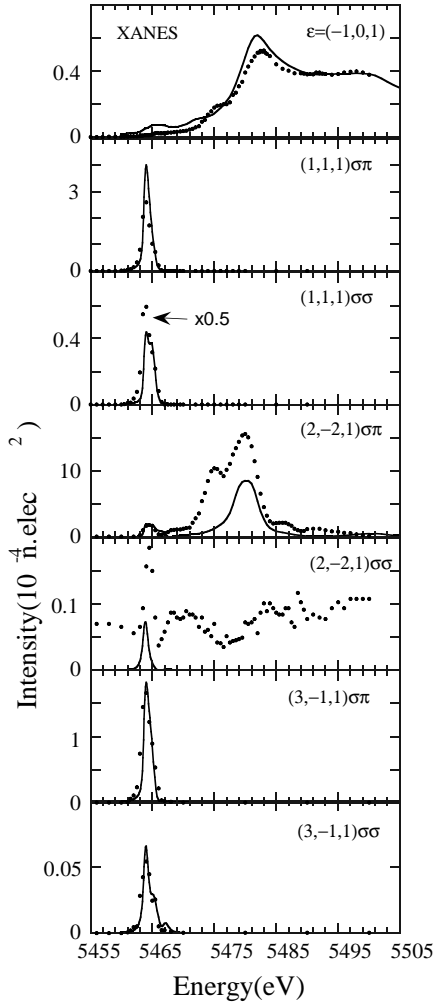,width=6cm}}
 \vspace{170pt}
 \caption {Comparison of the experimental (dots) and
calculated (continuous line) energy scans for three different
reflections. On top is also shown the XANES spectrum.
 The calculation is performed on a 4.3 {\AA} cluster radius
 and includes the magnetic ordering. The intensity unit is the electron 
number squared. Note that
the experimental spectra were recorded\cite{paolasini2}
with a special care on the relative amplitudes. For the
$(2,\bar{2},1)$ and $(3,\bar{1},1)$ reflections the azimuthal
angle as defined by Paolasini and coworkers is $\psi = 15^0$. All
the mean features are present. An extra scale factor is applied on
the experimental dots of the $(1,1,1)_{\sigma\sigma}$ reflection.} 
\label{figure4_spectre}
\end{figure}

\begin{figure}
\vspace{70pt}
 \centerline{\epsfig{file=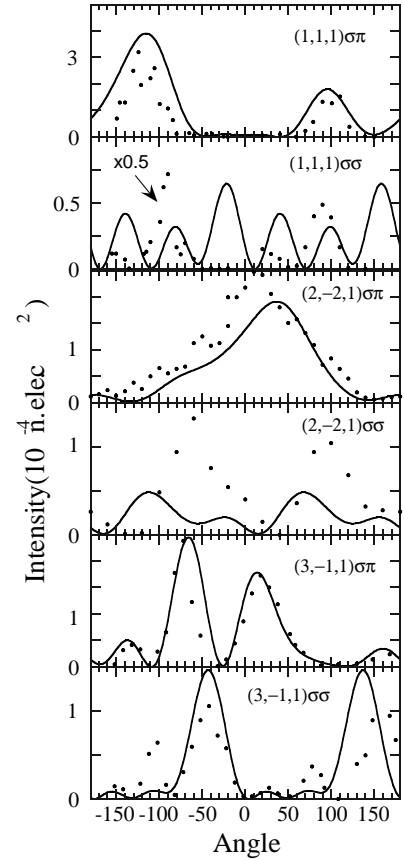,width=6cm}}
  \vspace{100pt}
 \caption {Azimuthal scans of the same three reflections shown in Fig. 
\ref{figure4_spectre}. The main
features of the experiment (dots) are obtained in the
calculation (continuous line). The scans are performed
respectively at 5464 eV for the $(1,1,1)$ and $(3,\bar{1},1)$
reflections and 5465 eV for the $(2,\bar{2},1)$. The intensity unit is the 
number of electrons squared.
The azimuth 90$^0$ corresponds to the situation where the scattering
plane contains the $c$ hexagonal axis. }
 \label{figure5_scan}
\end{figure}

The behaviour of the azimuthal scans against the cluster radius looks more
complex. The agreement improves up to the 4.3 {\AA}, corresponding to 33 
atoms, and then
decreases when the cluster radius is further increased. The reason for this
is discussed below. For the moment we keep the 4.3 {\AA}
radius and compare such energy and azimuthal spectra with the experimental 
ones, 
as shown in Figs. \ref{figure4_spectre} and \ref{figure5_scan}.

Looking at the spectra shown in Fig. \ref{figure4_spectre},
we can claim a satisfactory experiment-theory agreement for the
different reflections. In particular we get the very thin
resonance in the $3d-t_{2g}$ energy range at the $(1,1,1)$ and
$(3,\bar{1},1)$ peaks for both $\sigma\sigma$ and $\sigma\pi$
polarization conditions and for the $(2,\bar{2},1)_{\sigma\sigma}$
reflection. The broad intensity in the $4p$ energy range in the
$(2,\bar{2},1)_{\sigma\pi}$ is also reproduced. 

The intensity
ratio between  $\sigma\sigma$ and $\sigma\pi$ channels and
between the various reflections are also quite good considering
the difficulties on both experimental and theoretical sides,
even if for the $(1,1,1)_{\sigma\sigma}$ reflection an extra factor
improves the agreement with the experiment. Note that, differently from all 
other reflections, the $(2,\bar{2},1)_{\sigma\pi}$ can contain some 
non-resonant
component not taken into account in our calculation. Moreover, a non-zero 
offset seems to be present at $(2,\bar{2},1)_{\sigma\sigma}$ on the 
experimental side, and this is probably
responsible for the discrepancy in the corresponding energy and azimuthal
scans. The azimuthal scans, shown in Fig. \ref{figure5_scan}, are reasonable 
for all $\sigma\pi$ reflections. Only the $(1,1,1)_{\sigma\sigma}$ is much 
less satisfying: yet, the corresponding data belong to the first experiment 
(Ref. \onlinecite{paolasini1}, while the others belong to Ref. 
\onlinecite{paolasini2}), and this does not allow to be sure about the 
relative intensity as in the other cases. Finally, we are also able to 
reproduce the $\pi$ and $2\pi$ periodicity, expected from the low symmetry 
of the compound, for the $\sigma\sigma$ and $\sigma\pi$ channels, respectively.

\begin{figure}
\vspace{20pt}
 \centerline{\epsfig{file=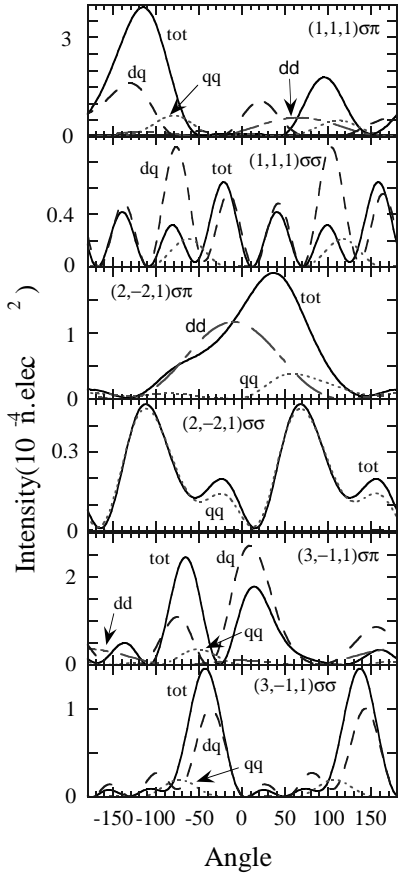,width=6cm}}
  \vspace{150pt}
 \caption {Magnetic intensity of the various reflections
at the vanadium K edge. The total
(continuous line), dd (dash-dotted line), dq (dash line) and qq
(dotted line) signals are shown, at the energy E=5465 eV} \label{figure6_dp}
\end{figure}

We judge the agreement between theory and experiment quite satisfactory. 
Indeed the interference between the dd, dq and qq parts makes the problem
really tricky. Foe example, $p$ orbitals are far more
delocalized than $d$ ones. For this reason they react in a different way 
to the core hole screening.
In a first approximation $d$ states are more
shifted towards lower energy than  $p$ states. Thus, the dq terms, which
probe the hybridization between $p$ and $d$ orbitals, are strongly influenced
by the details of the electronic structure around the
photo-absorbing ion. If we consider that
V$_2$O$_3$ is a strongly correlated electron system, with a quantum-entangled 
ground-state wave function,\cite{mila,dimatteo,tanaka} we find quite
surprising that we have reached such a good agreement in the $3d$ energy 
region with a one-electron calculation. Our idea is that this can help 
understanding which features can be explained in terms of an independent 
particle approach and which cannot (see below). Note that the localization 
of the $d$ orbitals also explains why the
azimuthal scans look better when the calculations are performed using the
smaller 4.3 {\AA} cluster radius.
To end this discussion we stress that energy spectra profiles around 
5464 eV are very sensitive to the azimuthal angle and, vice-versa, 
azimuthal profiles are very sensitive to the energy. This is due to 
the fact that qq and dq components have both strong (and different) 
angular and energy dependence. There are some angles where the calculated 
profiles appear doubled, with about 1 eV between the two peaks. For instance, 
this is what happens at the experimental azimuth of the 
$(3,\bar{1},1)_{\sigma\sigma}$ energy spectrum: a simple shoulder appears 
at 5465 eV, well described by our calculation.

Finally, we want to comment about the two papers (by Lovesey {\it et al.}
\cite{lovesey2} and by Tanaka.\cite{tanaka}) that have already pointed out 
the magnetic origin of the $h+k+l=$odd reflections. 
These works disagree on the physical mechanism of the signal, as for Tanaka 
only dq terms contribute, while Lovesey and coworkers attribute the signal 
to a pure qq reflection.
Our results on this point show that both dq and qq channels can contribute 
to the global intensity depending on the reflections (whether $h=$odd or 
$h=$even) and on the azimuthal angle (see Fig. \ref{figure6_dp}).  
At the prepeak energy, E=5465 eV, the dq term represents the strongest 
contribution for the $(1,1,1)$ and $(3,\bar{1},1)$ reflections,
even if the qq term is not completely  negligeable. For the 
$(2,\bar{2},1)_{\sigma\sigma}$, on the contrary, the qq contribution is 
clearly the dominating one. Finally, it is interesting to note that for the 
$(2,\bar{2},1)_{\sigma\pi}$ reflection the dd term plays the major 
role even at the $3d$ energies. Anyway the qq contribution has always a 
non-negligeable influence on the total signal.
Comparing our findings with those of Refs. \onlinecite{lovesey2,tanaka} 
we can state that there are no strict extinction rules regarding dq and 
qq terms for all these reflections. Nonetheless, the main contribution to 
$h+k+l=$odd, $h=$odd reflections comes from the dq channel (as in Ref. 
\onlinecite{tanaka}), while the main term in $h+k+l=$odd, $h=$even 
reflections is of qq origin (as in Ref. \onlinecite{lovesey2}).
Such a tensor analysis allows also to identify a feature that is not 
correclty described by our one-electron calculations, ie, the direction 
of the magnetic moment. Indeed, a small $L_y$ component (about $10\%$ the 
global moment) is found to contribute to the $h=$odd signal (see Fig. 
\ref{figure6_dp}), contrary to what expected by our previous theoretical 
analysis. This result can be explained by noticing that in order to obtain 
the correct direction of the magnetic moment a molecular, correlated 
ground-state wave function is needed,\cite{tanaka} which is far beyond 
the possibilities of our monoelectronic approach.

\subsection{Analysis of the "orbital" signal}

In this subsection, we analyze the effect of a time-reversal breaking OO 
on the signal, in order to determine whether it can affect our previous 
results. This is not a secondary issue since the original experimental 
interpretation suggested that the (111)-monoclinic reflection was due to 
the OO\cite{paolasini1} and a big controversy around this point arose in 
the literature.\cite{ezhov,mila,dimatteo,tanaka} The reason why we focus on 
a "time-reversal breaking" OO is that, for $h+k+l=$odd, the signal is 
proportional to $({\hat {E}}-{\hat {T}})f_1$, as clear from Eq. (\ref{eq_t}),  
so that a non-magnetic signal is allowed only when the time-reversal 
symmetry is broken. This is obtained, for example, when V$_i$ and V$_{i'}$ 
have a different orbital occupancy,\cite{dimatteo} and we call such a 
situation time-reversal breaking OO.
Different types of orbitally-ordered ground states have been suggested in 
the literature,\cite{mila,dimatteo,cnr} but none of them possesses such a 
feature. As a consequence, non-magnetic signals at the (111)$_m$ reflection 
are not expected, unless some {\it ad hoc} hypoteses are made, as recognized 
in the discussion of Sec. VII.D of Ref. [\onlinecite{dimatteo}]. One possible 
way to get a signal of OO origin is to consider the first excited state found 
in Ref. [\onlinecite{dimatteo}], which has a magnetoelectric (ME) symmetry. 
It is very close to the ground state ($\simeq1$ meV) and, because of this, 
it could be partly occupied. 
Independently of this, if we consider the two maximal ME subgroups of the 
full space groups, ie, P2'/a (with symmetry elements ${\hat E}$, 
${\hat T \hat I}$, ${\hat T \hat C_{2b}}$, ${\hat m_{b}}$) and P2/a' 
(with symmetry elements ${\hat E}$, ${\hat T \hat I}$, ${\hat C_{2b}}$, 
${\hat T \hat m_{b}}$),
it is in principle possible to get a non-magnetic signal at the (111)$_m$ 
reflection. In fact, in both cases there are two subgroups of four ions 
whose electronic densities have anisotropies not connected by time-reversal 
nor any other symmetry operation. 
Consider, for example, the case of P2/a', that could be responsible for 
non-reciprocal dichroism\cite{lindic}: the two groups of atoms 
($V_1, V_{2'}, V_3, V_{4'}$) and ($V_{1'}$, $V_{2}$, $V_{3'}$, $V_{4}$) 
are independent and the structure factor, Eq. \ref{eq_t}, becomes

\begin{eqnarray}
{\cal F}^{hkl}=(1+(-)^h\hat{T}\hat{m_b}) (e^{i\phi_{hkl}}+(-)^{h+k+l}
e^{-i\phi_{hkl}}\hat{T}\hat{I}) \nonumber \\
(f_1+(-)^{h+k+l}f_{1'})
\label{eq_t2}
\end{eqnarray}

The analogous of Eq. \ref{eq_d} for reflections with $h+k+l=$odd is now:

\begin{eqnarray}
{\cal D}_{\alpha\beta}^{hkl} = 4i(1+(-1)^{h+n_y})\sin(\phi_{hkl})
\Re(D_{\alpha\beta}) \nonumber \\
+4i(1+(-1)^{h+n_y+1})\cos(\phi_{hkl}) \Im(D_{\alpha\beta}) \nonumber \\
{\cal I}_{\alpha\beta\gamma}^{hkl} = 4(1+(-1)^{h+n_y})
\cos(\phi_{hkl})\Re(I_{\alpha\beta\gamma}) \nonumber \\
+4(1+(-1)^{h+n_y+1})\sin(\phi_{hkl})\Im(I_{\alpha\beta\gamma}) \nonumber \\
{\cal Q}_{\alpha\beta\gamma\delta}^{hkl} = 4i(1+(-1)^{h+n_y})
\sin(\phi_{hkl})\Re(Q_{\alpha\beta\gamma\delta}) \nonumber \\
4i(1+(-1)^{h+n_y+1})
\cos(\phi_{hkl})\Im(Q_{\alpha\beta\gamma\delta})
\label{eq_d2}
\end{eqnarray}

Thus, the breakdown of time-reversal symmetry, with the introduction of the 
two ASF $f_1$ and $f_{1'}$, has made possible also non-magnetic scattering, 
through the real part of the tensor components. Notice that, due to the 
magnetic space group, both magnetic and non-magnetic scattering can in 
principle interfere at particular polarization conditions and reflections.

In order to calculate the signal explicitly, we need to express the orbital 
wave functions of V$_1$ and V$_{1'}$ ions. Starting from the orbital 
occupancy of Refs. \onlinecite{mila,dimatteo}, we get, in the reference 
frame of Fig. \ref{fig_1}:

\begin{eqnarray*}
\phi_1 = \frac{1}{3}d_{xy}
       + \frac{\sqrt{2}}{3}d_{yz} + \frac{1}{\sqrt{6}}d_{z^2}
       - \frac{1}{3\sqrt{2}}d_{xz}
       - \frac{2}{3}d_{x^2-y^2} \\
\phi_{1'} = \frac{2}{3}d_{xy}
       + \frac{1}{3\sqrt{2}}d_{yz} + \frac{1}{\sqrt{6}}d_{z^2}
       - \frac{\sqrt{2}}{3}d_{xz}
       - \frac{1}{3}d_{x^2-y^2}
\end{eqnarray*}
\vspace{-1cm}
\begin{equation}
\label{eq_phi}
\end{equation}

From this orbital occupancy, we can calculate the electron density.
Then, solving Poisson's equation, we get the Coulomb potential. The
energy-dependent exchange-correlation potential is then obtained with
conventional procedures.\cite{yvestio2}
The calculation of the ASF is performed for the $V_1$ and $V_{1'}$ atoms 
with the FDM option. We kept a rather small cluster radius (3.0 {\AA}), 
but included the oxygen octahedra as well as the first shell of four vanadium 
neighbors. The ASF of the other atoms are then deduced by means of the 
magnetic symmetries of Table I for the two separate subgroups 
(see also Ref. [\onlinecite{dimatteo}], sec. VII.D, for a derivation) and 
the total scattering amplitude is obtained using Eq. (\ref{eq_f}).

\begin{figure}
\vspace{-20pt}
\centerline{\epsfig{file=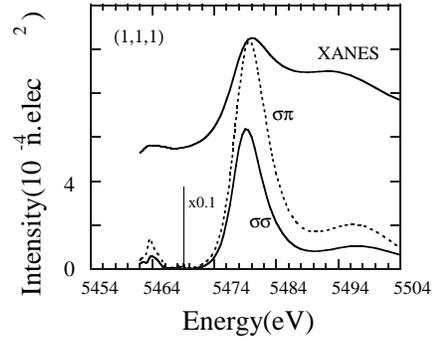,width=6cm}}
\vspace{-10pt}
\caption{Intensity of the $(1,1,1)_{\sigma\sigma}$ (full line) and
$(1,1,1)_{\sigma\pi}$ (dotted line) reflections at the vanadium K edge 
calculated with the orbital ordering of Eq. (\ref{eq_phi}). The Bragg
peaks have the same energy scale as the XANES spectra (full line). The big
structure at the 4p edge is not present in the experiment.
The calculation is performed with a 3 {\AA} cluster radius.}
\label{fig_2}
\end{figure}

Our results for the $(1,1,1)_m$ reflection are shown in  Fig. \ref{fig_2}.
As in the experiment,\cite{paolasini1} a structure is obtained at the 
3d-energy level. Yet, a much bigger signal is present
at the 4p edge where the experimental intensity shows no features.
Such a disagreement is the proof that the OO cannot be responsible
of the signal: The time-reversal breaking OO makes the $(111)_m$ 
reflection allowed not only
at the 3d-energy level, but also at the 4p-energies. Note that this result
remains valid for any time-reversal breaking OO, not only the one proposed 
in Eq. (\ref{eq_phi}), as should have been qualitatively expected. Indeed, it 
has already been shown\cite{benfatto,elfimov} that the 3d-4p hybridization, 
even if small, must give a contribution to the dipolar K edge signal. 
In fact, previous numerical simulations in another compound, LaMnO$_3$, have 
shown that the effect of a symmetry-breaking OO of 3d orbitals produces a 
signal also at the 4p-energy levels. In this latter case the contribution 
coming from the Jahn-Teller distortion overwhelms the OO signal. However 
in the present case, since V$_1$ and V$_{1'}$ have the same local distortion, 
such a signal should have been detected, as clear from our numerical 
simulation shown in Fig. \ref{fig_2}. Thus, the experimental evidence 
of absence of any signal at the 4p energies proves two facts: first, there 
is no time-reversal breaking OO; second, the (111)$_m$ cannot be due to OO. 
A direct consequence of the previous analysis is that no magnetoelectric s
ubgroups are compatible with the (111)$_m$ energy scan, as they all break 
the time-reversal symmetry, in keeping with the negative experimental 
evidence for magnetoelectricity in V$_2$O$_3$.\cite{jansen} This strongly 
supports the results of Ref. [\onlinecite{lindic}].

Notice that a previous study\cite{cuozzo} on the same
reflection in V$_2$O$_3$ ended up with the conclusion that the OO was
responsible of the signal. Unfortunately that
calculation was performed only around the 3d-energy range, where the 
agreement in the azimuthal scan was quite good and the big feature at 
the 4p energies was not detected. 
The lesson to be drawn from this fact is that the azimuthal scan
around the diffraction vector is not always a fundamental feature in 
determining the origin of the reflection.
In fact, it usually reflects more the geometry than the
electronic properties of the material. 
This is a rather general comment, not limited to V$_2$O$_3$: when there are 
only one or two {\it ab
initio} independent factors (the tensor components), the angular properties 
are much more determined by the geometry (ie, the symmetries) than by the 
dynamics (ie, the relative weight of the radial matrix elements). In this 
situation only the combined analysis of both energy and azimuthal scans is 
a reliable tool to investigate the electronic origin of RXS phenomena.

\section{Conclusion}

The main results of the present paper can be inferred from the conclusions 
of the last two subsections discussing the results appeared in the 
literature. In fact, up to now, it was possible to classify the 
interpretations of the forbidden Bragg reflections with $h+k+l=$odd and 
$h=$odd along two main lines of thought: those who explained such 
reflections in terms of orbital ordering, both associated\cite{dimatteo} 
or not\cite{mila} with a reduction of the magnetic symmetry, and those who 
were inclined to a magnetic origin.\cite{lovesey2,tanaka}
A third explanation, that of the antiferroquadrupolar ordering 
proposed in Ref. [\onlinecite{ezhov}] had already been ruled out in Ref. 
[\onlinecite{dimatteo}]. Whether the former or the latter interpretations 
were correct, was not simple to decide on the basis of the azimuthal scans, 
only. In fact, such scans, for a given multipolar channel, measure
the crystal symmetry, rather than the electronic origin of the reflections. 
The proof of such a statement is that three different mechanisms (magnetic 
in the qq channel\cite{lovesey2}, magnetic in the dq channel\cite{tanaka} 
and non-magnetic, via orbital ordering\cite{cuozzo}) gave all a rather 
good agreement in the azimuthal scan at the pre-K edge.
What definitively rules out the OO origin of such
reflections is the energy scan shown in Fig. 6. As noted previously, a 
direct consequence of this result is that no reduction of the magnetic 
space group $P2/a+{\hat {T}}\{{\hat{E}}|t_0\}P2/a$  is present.
\cite{lindic,jansen}

The second important result of the present paper lies in the fact that we 
were able to make a complete {\it ab-initio} analysis of all $h+k+l=$odd 
reflections, starting from the crystal and magnetic structure, only. 
We showed that their origin (whether dd, dq or qq) strongly depends on the 
kind of reflection (ie, $h=$even or $h=$odd), photon energy and azimuthal 
angle. For  $h=$odd, the dq channel is predominant, in 
keeping with the cluster calculations of Tanaka.\cite{tanaka}
In spite of this particular agreement, we believe that it is important to 
stress that our approach goes beyond the cluster calculations performed in Ref.
[\onlinecite{tanaka}] as well as the simple fitting procedure of Ref.
[\onlinecite{lovesey2}]. In fact only by means of an {\it ab-initio} 
procedure it is possible to cover all the main experimental evidence, ie, 
the azimuthal behaviour, the energy profiles, and the order of magnitude 
of the intensity, that is well in keeping with the rough estimate given in 
Ref. [\onlinecite{paolasini1}].  

We would like to acknowledge L. Paolasini for several useful discussions.

\end{multicols}


\begin{thebibliography}{99}
\bibitem{note1} The magnetic dipole term can be safely neglected in the 
x-ray range
\bibitem{nota0}
The fact that ${\cal {D}}_{\alpha\beta}-{\cal {D}}_{\beta\alpha}$ is 
proportional to the angular orbital moment can be shown as follows. 
Using the identity:
\begin{eqnarray*}
{\cal {D}}^{hkl}_{xy}-{\cal {D}}^{hkl}_{yx}\propto \sum_{gn} 
|\langle \psi_n |x+iy | \psi_g \rangle |^2-
|\langle \psi_n |x-iy | \psi_g \rangle |^2
\end{eqnarray*}

\vspace{-0.4cm}

\noindent the right hand side is the usual expression for magnetic circular
dichroism in the dipole channel, which can be shown \cite{dimatoli} to be
proportional to $\langle L_z \rangle$ at the K-edge.
\end{thebibliography}
\end{document}